\documentclass[review]{elsarticle}
\usepackage{amsmath, mathrsfs, amssymb,amsfonts,amsthm,graphicx, epsf, dcolumn}
\usepackage{mathtools}
\usepackage[hyperfootnotes=true]{hyperref}
\usepackage{subfigure}
\usepackage{color}
\usepackage{slashed}
\usepackage{setspace}
\usepackage{cancel}
\usepackage{wasysym}
\usepackage{hyperref}
\usepackage{float}
\usepackage{soul}

\usepackage[normalem]{ulem}
\definecolor{orange}{rgb}{1,0.5,0}
\definecolor{pink}{RGB}{255,0,255}
\definecolor{purple}{RGB}{128,0,128}

\pdfoutput=1
\newcommand{\be}{\begin{equation}}
\newcommand{\ee}{\end{equation}}
\newcommand{\bea}{\begin{eqnarray}}
\newcommand{\eea}{\end{eqnarray}}

\begin{document}
\date{}
\begin{frontmatter}

\title{Inflationary predictions of Geometric Inflation}

\author[1]{Gustavo Arciniega\corref{cor1}\fnref{fn1}
}
\ead{gustavo.arciniega@ciencias.unam.mx}

\author[2]{Luisa G. Jaime\fnref{fn1}
}
\ead{luisa.jaime1@gmail.com}

\author[1]{Gabriella Piccinelli\fnref{fn2}}
\ead{itzamna@unam.mx}

\cortext[cor1]{Corresponding author}
\fntext[fn1]{Present address}
\fntext[fn2]{Permanent address}

\address[1]{Centro Tecnol\'ogico, Facultad de Estudios Superiores Arag\'on, Universidad Nacional Aut\'onoma de M\'exico. Avenida Rancho Seco S/N Col. Impulsora Popular Av\'icola, Nezahualcoyolt, Edo. de M\'exico 57130, Mexico.}
\address[2]{Departamento de F\'isica, Instituto Nacional de Investigaciones Nucleares, Apartado Postal 18-1027, Col. Escand\'on, Ciudad de M\'exico,11801, Mexico.}


\begin{abstract}
In the framework of gravitational models obtained from the Geometric Inflation's proposal, where an infinite tower of curvature scalars are included into the action, we compute the slow-roll parameters by the Hubble slow-roll approach. We test the viability of such models as inflationary scenarios, focusing on the tensor-to-scalar ratio, $r$, and spectral scalar index, $n_s$, relation. We find that all models considered here produce inflation and, most of them coincide, some better than others, with the marginalized 95$\%$ CL region given by Planck's data collaboration. 
\end{abstract}


\end{frontmatter}

\section{Introduction}
\label{secc:Introduction}

Currently, cosmic inflation is widely accepted as a necessary mechanism at the beginning of the Universe \cite{Guth:1980zm, Linde:1981mu, Baumann:2009ds} for solving some cosmological puzzles, as the horizon and flatness problems. This scenario is, besides, well supported by cosmological observations \cite{Akrami:2018odb}. For achieving an inflationary process, the strong energy condition ($\rho + 3p \geq 0$) does not have to be satisfied and this can be accomplished in two ways: invoking some kind of field with negative pressure (see \cite{Martin:2013tda} for a review and references therein), which is the standard approach, or introducing some modification to the gravitational sector \cite{Lyth:2009zz, Clifton:2011jh}.

It is a well-known fact that the inclusion of higher curvature corrections can modify the initial expansion of the Universe. The most relevant example for this is the addition to the General Relativity (GR) action of a $R^2$ term, performed by Starobinsky \cite{Starobinsky:1980te}. It is worth to mention that observations of the ratio of tensor to scalar modes, $r$, seem to be in a good agreement with the predictions of Starobinsky inflation \cite{Akrami:2018odb}, nevertheless, the transition to late Universe is not well supported under such a model \cite{Amendola:2006kh, Jaime:2012yi}.

In \cite{Arciniega:2018fxj}, a cubic term in the curvature invariants is added as a correction to GR for cosmology, soon after a quartic and quintic term is considered in \cite{Cisterna:2018tgx}, and in \cite{Arciniega:2018tnn, Arciniega:2019oxa} the theory is generalized to an infinite tower of curvature scalars.  This addition shares some nice properties with GR, as the existence of Schwarzschild-like solutions and no-hairy black holes, degrees of freedom propagating (in a vacuum) that correspond to the standard graviton, and, moreover, a well-behaved cosmology as an initial value problem with second-order Friedmann equations for a homogeneous and isotropic universe. 

A practical way of studying the evolution of the inflationary process is through the analysis of the slow-roll regime. Since the beginning of the 2000s \cite{Schwarz:2001vv, Leach:2002ar}, it has been shown that the cosmological perturbation parameters (tensor-to-scalar ratio, $r$, spectral indexes, $n_s$ and $n_T$, and their runnings, $\alpha_s$ and $\alpha_T$) can be calculated through the horizon-flow parameters, $\epsilon_n$ (also called Hubble slow-roll parameters), instead of the slow-roll approach given through a potential $V$ \cite{Lyth:2009zz, Steinhardt:1984jj}. This has the advantage that $\epsilon_{n}$ are model-independent parameters in the sense that they have a dependency on $H$ and its derivatives only.

The relation of the cosmological observables with the $\epsilon_n$ parameters has been obtained for inflationary models with a potential scalar field associated with the model. Even though the scenario presented in \cite{Arciniega:2018tnn} does not have an explicit scalar potential $V$, it is plausible to consider that those equalities hold, as happens with quadratic curvature invariant models such as \cite{Starobinsky:1980te}. Moreover, in recent works on string theory where a scalar potential $V$ is present \cite{Hohm:2019omv, Hohm:2019jgu}, field equations similar to the ones shown in \cite{Arciniega:2018tnn} have been obtained when an infinitely higher-derivative $\alpha'$ corrections (with $\alpha'$ the coupling constant of the string) are taken into account for the string scenario in cosmology.  In this case, the computation of the cosmological observables can be obtained through the slow-roll approach.

In the present paper, we further explore a geometric modification to GR where more curvature scalars, besides $R$, are taken into account when the action is proposed. We analyze the inflationary epoch in the slow-roll approximation, using the horizon-flow parameters for six cases in the framework of this theory, where an initial exponential acceleration is predicted and then attenuated, as the scalar factor, $a(t)$, grows, in such a way that the graceful-exit problem is solved and the evolution to late Universe, where $\Lambda$ dominates, is naturally reached.  

\section{The theory}
\label{TheTheory}

We will consider the action given in \cite{Arciniega:2018tnn}:

\begin{equation}\label{theo}
S=\int \frac{d^4x \sqrt{|g|}}{16\pi G}\left\{-2\Lambda+R+\sum_{n=3}^{\infty}\lambda_n L^{2n-2}\mathcal{R}_{(n)}\right\}\, ,
\end{equation}

where $R + \Lambda$ is the standard action for General Relativity plus cosmological constant, $\mathcal{R}_{(n)}$ are Lagrangian densities constructed by curvature scalars upto $n$th-order\footnote{For the construction of the Lagrangian densities considered in this work, see Appendix 1 in \cite{Arciniega:2018tnn}.}, $\lambda_n$ are dimensionless couplings, and $L^{-1}$ is playing the role of a new energy scale.


\subsection{Properties of the theory}

The equations of motion coming from this action have the following nice properties:
\begin{enumerate}[i)]
\item Second order linearized equations on maximally symmetric backgrounds \citep{Bueno:2016ypa}.
\item Single-function Schwarzschild-like solutions and no-hairy black holes \cite{Bueno:2017sui}.
\item Graceful exit from their inflationary epoch \cite{Arciniega:2018tnn}.
\item Well-posed initial value problem for an homogeneous and isotropic Universe \cite{Arciniega:2018tnn}.
\end{enumerate}

The analysis regarding the well-posed cosmology was first introduced in \cite{Arciniega:2018fxj} for the cubic case, and in \cite{Cisterna:2018tgx} for the quartic and quintic case.

In \cite{Bueno:2016ypa}, a linearization procedure for Lagrangian densities, $\mathcal{L}$(Riemann) theories, for maximally symmetric backgrounds is developed. In general, at linear order these theories contain a ghost-like massive spin-2, $m_g$, and a scalar mode, $m_s$.   Under the linearization procedure it is possible to impose algebraic conditions that guarantee that $m_g$ and $m_s$ modes do not propagate (Einsteinian-like theories). Modifications of gravity could have different kinds of problems at second or even at first order in perturbations. The present theory has the advantage to avoid problems at first order. Second-order cosmological perturbations are something that should be explored, particularly for predictions at the late-time universe. 

\subsection{Modified Friedmann equations}

In order to explore the cosmic evolution in the framework of the present work, we use the standard Friedmann-Lema\^{i}tre-Robertson-Walker (FLRW) metric
\begin{equation}\label{FLRW}
ds^2=-dt^2+a(t)^2 \left(\frac{dr^2}{1-k r^2}+r^2 d\Omega^2 \right)\, ,
\end{equation}
where $a(t)$ is the scale factor. The modified Friedmann equations corresponding to the $t-t$ and $r-r$ components are respectively
\begin{align}\label{Fried}
3F(H)&=8\pi G \rho+\Lambda\, ,\\ \label{Fried2}
-\frac{\dot{H}}{H}F'(H)&=8 \pi G (\rho+P) \, ,
\end{align}
where we had considered a flat spatial curvature, $k=0$, and
\begin{equation}\label{F}
F(H)\equiv H^2+L^{-2}\sum_{n=3}^{\infty} (-1)^n\lambda_n \left(LH\right)^{2n}\, ,
\end{equation}
with $F'(H)\equiv dF(H)/dH$. $H=\dot{a}/a$ is the Hubble parameter, and $\rho$ and $P$ correspond to the density and pressure for a perfect fluid. Notice that if we set all the couplings $\lambda_n=0$, then  $F(H)=(\dot a/a)^2$, $\dot{H} F'(H)/H=2 (\ddot a a -\dot a^2)/a^2$ and the Friedmann equations for GR are recovered.

\section{Slow-roll parameters for Geometric Inflation}

As exposed in \cite{Schwarz:2001vv, Leach:2002ar}, the background evolution can be described by the horizon-flow parameters (HFP), $\epsilon_n$, also called Hubble slow-roll parameters (HSR) or Hubble flow functions (HFFs) \cite{Akrami:2018odb}, defined by the expression
\begin{equation}\label{epsilons}
\epsilon_{n+1}\equiv -\frac{d\,\mathrm{ln}|\epsilon_n|}{dN}=\frac{\dot{\epsilon}_n}{H \epsilon_n},
\end{equation}

\noindent where $\epsilon_0\equiv H(N_i)/H(N)$, and  $N\equiv \mathrm{ln}(a/a_i)$ is the e-fold number since some initial time $t_i$.

For the inflationary epoch, it is customary to neglect the $\Lambda$ term, and consider only a radiation domination matter content, i.e., $\rho=\rho_r$ and $P=(1/3)\rho_r$. From this, the generalized FLRW equations, (\ref{Fried}) and (\ref{Fried2}),  and equation (\ref{epsilons}), we can rewrite the first three HSR parameters in terms of $F(H)$ and its derivatives as
\begin{eqnarray}\label{hsr1}
&& \epsilon_1 = \frac{4F(H)}{H F'(H)},\\ \label{hsr2}
&& \epsilon_2 = \epsilon_1-4\left(\frac{F(H)}{F'(H)}\right)',\\ \label{hsr3}
&& \epsilon_3 = \epsilon_1\left[1+\frac{4H}{\epsilon_2}\left(\frac{F(H)}{F'(H)}\right)''\right].
\end{eqnarray}

Now it is possible to compute the cosmological observables tensor-to-scalar ratio $r$, the spectral indexes $n_{\text{S}}$ and $n_{\text{T}}$, and their running $\alpha_{\text{S}}$ and $\alpha_{\text{T}}$  using the following expressions at second-order in $\epsilon_n$ \cite{Leach:2002ar, Planck:2013jfk}:
\begin{eqnarray} \label{observables:r}
&& r=16 \epsilon_1,\\ \label{observables:ns}
&& n_{\text{S}} = 1-2 \epsilon_1-\epsilon_2-2\epsilon_1^2-(2 C+3)\epsilon_1\epsilon_2-C\epsilon_2\epsilon_3 ,
\\ \label{observables:nT}
&& n_{\text{T}} = -2\epsilon_1-2\epsilon_1^2-2(C+1)\epsilon_1\epsilon_2 ,\\ \label{observables:as}
&& \alpha_{\text{S}} = -2\epsilon_1\epsilon_2-\epsilon_2\epsilon_3,\\ \label{observables:aT}
&& \alpha_{\text{T}} = -2\epsilon_1\epsilon_2,
\end{eqnarray}

\noindent where $C=\text{ln}(2)+\gamma_{\text{E}}-2\approx -0.7296$, and $\gamma_{\text{E}}$ is the Euler-Mascheroni constant.

\section{The models.}

The models we are taking into account in the present paper are obtained under the  method introduced in \cite{Arciniega:2018tnn}. For an inflationary exponential $F(H)$ model, it is convenient to rewrite and split equation (\ref{F}) into an even and an odd part for the $(-1)^{n}$ factor, as follows:
\begin{equation}\label{foddeven}
F(H)= H^2+\frac{\lambda}{L^{2}}\sum^{\infty}_{n=0}\lambda^{\text{even}}_n (-L^2H^2)^{2n+4}+\frac{\lambda}{L^{2}}\sum^{\infty}_{n=0}\lambda^{\text{odd}}_n (-L^2H^2)^{2n+3},
\end{equation}

\noindent where we have taken the constant, $\lambda_n$, in the Lagrangian density of the action in (\ref{theo}), as the product of $\lambda_n^{\text{(odd / even)}}$ (a dimensionless parameter) with $\lambda$ (a constant with proper physical units) \textit{i.e.} $\lambda_n\rightarrow (\lambda)(\lambda_n^{\text{(odd, even)}})$.

Now, from the last expression, it is not difficult to choose conveniently  the $\lambda^{\text{even}}_n$, and $\lambda^{\text{odd}}_n$ coefficients for all $n=0,\ldots,\infty$, in order to obtain exponential-like models. In the present work, we have chosen the six models presented in Table 1.

\begin{table} 
\begin{center}
{\renewcommand{\arraystretch}{2} 
\begin{tabular}{|c|l|c|c|} 

\hline
Model & $F(H)$ & $\lambda^{\text{odd}}_n$ & $\lambda^{\text{even}}_n$\\
\hline
\color{red}{\bf{1)}} & $H^2+\lambda H^6 L^4\, e^{(HL)^2} $ & $-\frac{1}{(2n)!}$ & $\frac{1}{(2n+1)!}$  \\
\hline
\color{blue}{\bf{2)}} & $H^2+\lambda H^8 L^6\,  e^{(HL)^4} $ & $0$ & $\frac{1}{n!}$ \\
\hline
\color{green}{\bf{3)}} & $H^2+\lambda H^6 L^4\,  e^{(HL)^6} $ *& $-\frac{1}{(2n)!}$ & $\frac{1}{(2n+1)!}$  \\
\hline
\color{orange}{\bf{4)}} & $H^2+\lambda H^6 L^4\,  e^{(HL)^8} $ ** & $-\frac{1}{n!}$ &  $0$ \\
\hline
\color{pink}{\bf{5)}} & $H^2+\lambda H^6 L^4\,  e^{(HL)^{10}} $ *** & $-\frac{1}{(2n)!}$ & $\frac{1}{(2n+1)!}$  \\
\hline
\color{purple}{\bf{6)}} & $H^2+\lambda H^{12} L^{10}\,  e^{(HL)^{12}} $ * & $0$ & $\frac{1}{n!}$  \\ 
\hline
\end{tabular}
}
\caption{Models of Geometric Inflation. Column 1 assigns a number to each model, column 2 shows the form of the $F(H)$ function when values for $\lambda_n$ odd and even (columns 3 and 4 respectively) are taken. In order to obtain de desired values of the power on the exponential, it is necessary to take equal to zero some of the $\lambda_{p}(LH)^{p}$ terms in equation (\ref{foddeven}). In doing so, we had taken: * \,\, $F(H)=H^2+\frac{\lambda}{L^{2}}\sum^{\infty}_{n=0}\lambda^{\text{even}}_n (-L^2H^2)^{2(3n+1)+4}+\frac{\lambda}{L^{2}}\sum^{\infty}_{n=0}\lambda^{\text{odd}}_n (-L^2H^2)^{2(3n)+3}$. 
** \, $F(H)=H^2+\frac{\lambda}{L^{2}}\sum^{\infty}_{n=0}\lambda^{\text{even}}_n (-L^2H^2)^{2(2n+1)+4}+\frac{\lambda}{L^{2}}\sum^{\infty}_{n=0}\lambda^{\text{odd}}_n (-L^2H^2)^{2(2n)+3}$. 
*** $F(H)=H^2+\frac{\lambda}{L^{2}}\sum^{\infty}_{n=0}\lambda^{\text{even}}_n (-L^2H^2)^{2(5n+2)+4}+\frac{\lambda}{L^{2}}\sum^{\infty}_{n=0}\lambda^{\text{odd}}_n (-L^2H^2)^{2(5n)+3}$.}
\label{Table:Models}
\end{center} 
\end{table}

The models were selected under the criteria of simplicity and proximity to the behavior of the standard cosmological model (with $H$ approximately constant during the inflationary regime), omitting, in the present work, more elaborate models that can be easily constructed from equation (\ref{foddeven}).

\section{Slow-roll results}

We are interested here in testing the inflationary viability of models that belong to the modified theory given in \cite{Arciniega:2018tnn}.  Although we are considering particular models, that follow the above criteria, it is interesting to notice that any model constructed from (\ref{foddeven}) can be read as $F(H)=H^2+\lambda f(H),$ where $f(H)$ does not contain any $\lambda$ dependence. 
The former, combined with the fact that the HSR parameters are functions of the quotient $F(H)/F'(H)$, leads to $\epsilon_n$ parameters that do not depend on $\lambda$ when $N$ is large since the exponential term dominates over the quadratic one for large $H$. This is an unexpected nice property of the theory.

In order to analyze the standard inflationary parameters ($r,$ $n_s,$ $n_T,$ $\alpha_s,$ and $\alpha_T$) we perform the computation of the HSR parameters, $\epsilon_i$, according to eqs. (\ref{hsr1})$-$(\ref{hsr3}), focusing at the $e-$foldings from $N=50$ to $N=60$, and compare our theoretical predictions with the values given by the Planck collaboration.

\begin{figure}[H]
\includegraphics[scale=0.23]{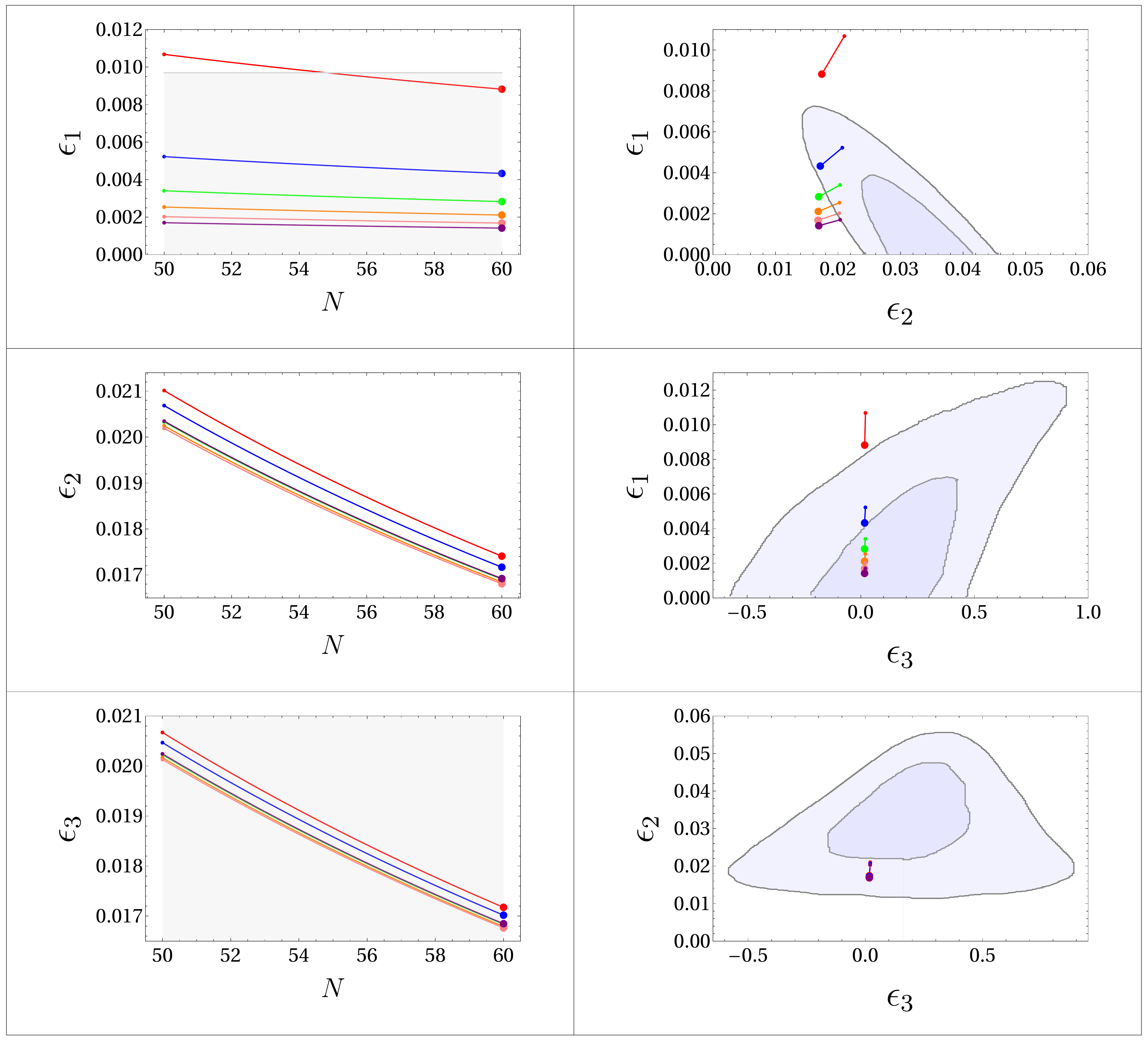} 
\caption{(Left) Evolution of $\epsilon_1$, $\epsilon_2$ and $\epsilon_3$ vs $N$ for models (from top to bottom: $1$-red, $2$-blue, $3$-green, $4$-orange, $5$-pink, and $6$-purple). According to Planck 2018 (shaded regions), the constraints are $\epsilon_1<0.0097$ (95$\%$ CL), $\epsilon_2=0.032(^{+0.009}_{-0.008})$ (68$\%$ CL) and $\epsilon_3=0.19(^{+0.55}_{-0.53})$ (95$\%$ CL). (Right) Combinations ($\epsilon_1$,  $\epsilon_2$, $\epsilon_3$) predictions for models $1-6$ with the $1$ and $2\sigma$ confidence level regions reported by Planck Collaboration. Plot $\epsilon_2-\epsilon_3$ seems overlapped for all models because the values for $\epsilon_2$ and $\epsilon_3$ are pretty similar, as is shown at left.}
\label{fig:epsilonPlanck}
\end{figure}

Figure \ref{fig:epsilonPlanck} shows, in the right side, the evolution of $\epsilon_i$ from $N=50$ to $60$ for the six models (Top to bottom: $1$-red, $2$-blue, $3$-green, $4$-orange, $5$-pink, and $6$-purple). In the case of $\epsilon_1$, all models are within the constriction $\epsilon_1<0.0097$ given by Planck (marginally for model $1$) \cite{Akrami:2018odb}. For $\epsilon_2$ all the models are below the inferior bound reported by Planck at marginalized $68\%$: $\epsilon_2 = 0.032( ^{+ 0.009}_{- 0.008})$, while in the case of $\epsilon_3$, which is very loosely constrained, they are localized in the $95\%$ CL zone. In the left side the combined predictions for $\epsilon_1 - \epsilon_2$ (top), $\epsilon_1 - \epsilon_3$ (middle) and $\epsilon_2 - \epsilon_3$ (bottom) with the marginalized $68\%$ and $95\%$ CL regions reported by Planck 2018 \cite{Akrami:2018odb} are plotted. In the $\epsilon_1-\epsilon_2$ space, we can clearly notice that model 1 is out of the $2\sigma$ region, model 2 is fully in the $2\sigma$ region while the other models reach, marginally, the $2\sigma$ contour. In the $\epsilon_1-\epsilon_2$ space, model 1 is out of the CL's. Model 2 is also in the $2\sigma$ region and the other models are in the $1\sigma$ contour. Observational errors for $\epsilon_3$ are still very large, nevertheless it is worth to notice that the predictions for $\epsilon_3$ for all the models explored are located in the $95\%$ CL region, $\epsilon_3=0.19(^{+0.55}_{-0.53})$, reported by \cite{Akrami:2018odb}. In Fig. \ref{fig:epsilonPlanck} bottom-right, all the models overlap, since $\epsilon_2$ and $\epsilon_3$ are almost equal, as can be seen in Fig. \ref{fig:epsilonPlanck} middle-left and bottom-left.

\begin{figure}
\includegraphics[scale=0.34]{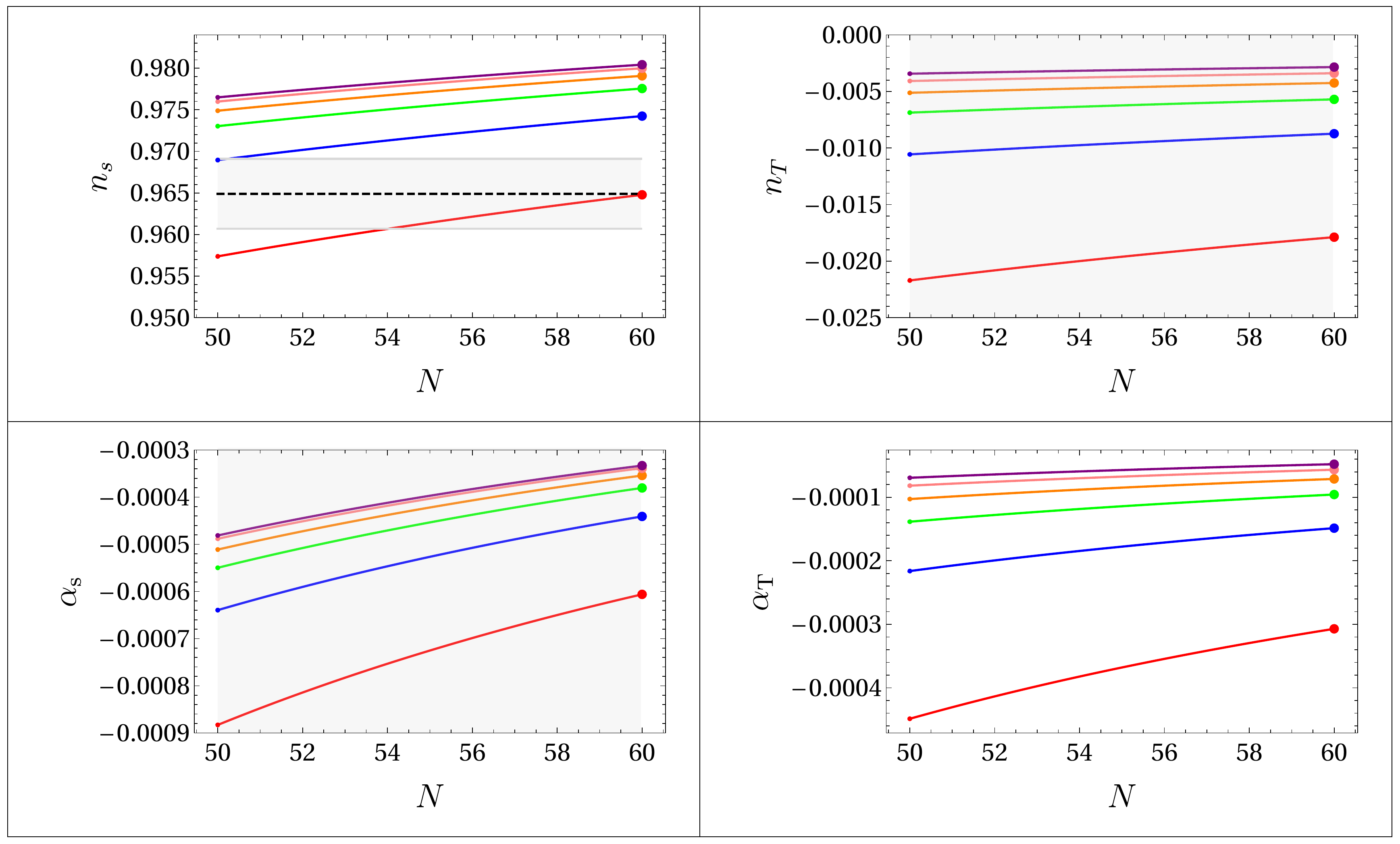}
\caption{Evolution of $n_s$, $n_T$, $\alpha_s$, and $\alpha_T$ for the models presented in Table (\ref{Table:Models}) (bottom to top: $1$-red, $2$-blue, $3$-green, $4$-orange, $5$-pink, and $6$-purple) from $N=50$ to $N=60$. The horizontal dashed line and shaded regions are, respectively, the central value and uncertainties given by Planck 2018 (68$\%$ CL). There is no reported value by Planck for $\alpha_T$.}
\label{fig:alphasN}
\end{figure}

\begin{figure}
\includegraphics[scale=0.7]{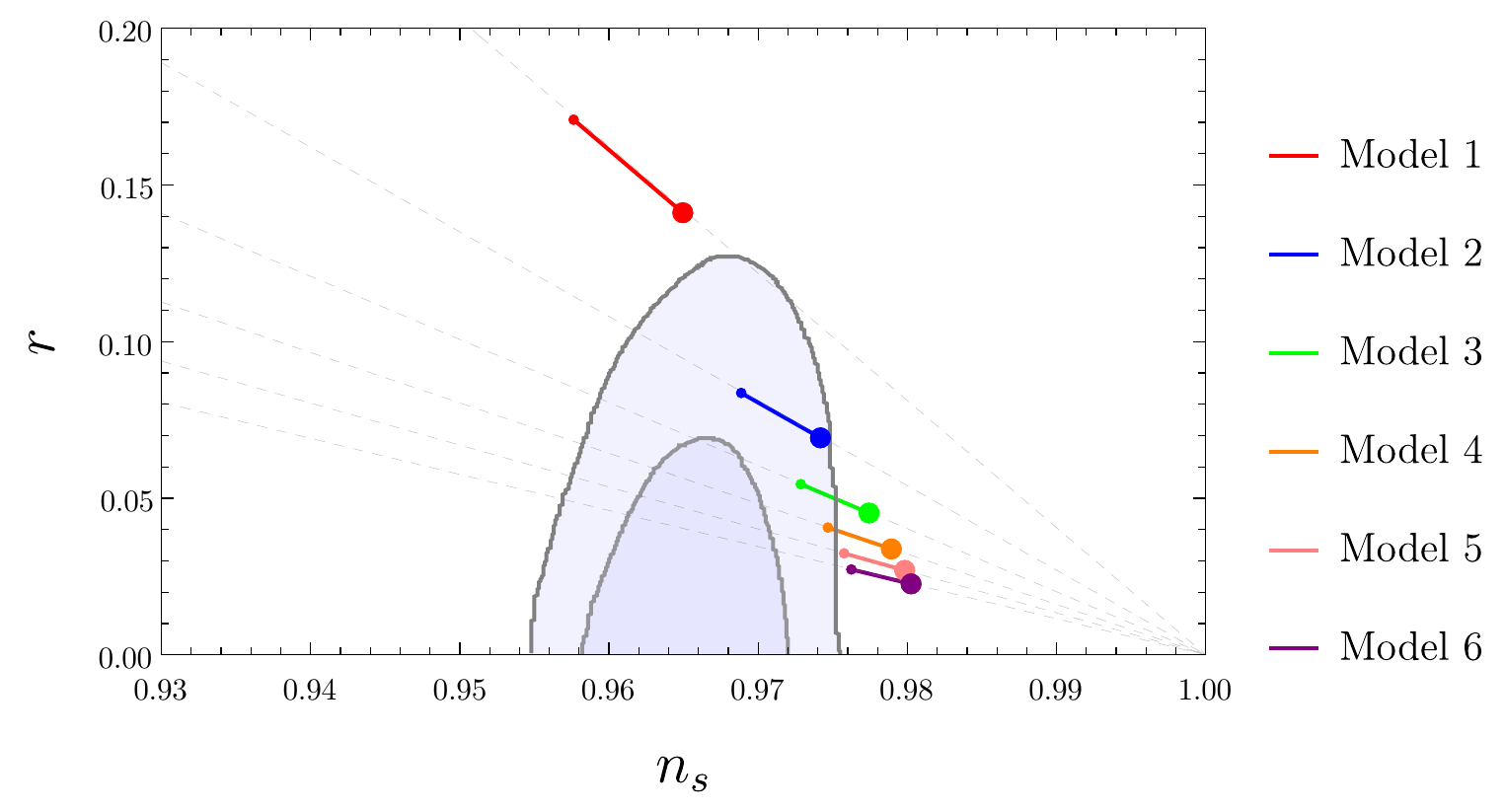}
\caption{Predicted $r$ vs. $n_s$ for models $1$ to $6$ (top to bottom: $1$-red, $2$-blue, $3$-green, $4$-orange, $5$-pink, and $6$-purple). The segments show the evolution from $N=60$ (big dot) to $N=50$ (small dot). The shadowed region shows the marginalized $68\%$ and $95\%$ CL regions from Planck 2018.}
\label{fig:nsvsr}
\end{figure}

According to eqs (\ref{observables:r})$-$(\ref{observables:nT}) we compute the predictions for the six models in Table (\ref{Table:Models}) for $r$, $n_s$, $n_T$, $\alpha_s$, and $\alpha_T$. Figure (\ref{fig:alphasN}) shows the evolution for each of these parameters from $N=50$ to $N=60$. Finally, we plot the tensor-to-scalar ratio $r$ and the spectral index $n_s$ in figure \ref{fig:nsvsr}, we can notice model $1$, $5$, and $6$ are out of the $95\%$ CL regions while model $2$ is fully contained in the $95\%$ CL and models $3$ and $4$ reach marginally this contour.

It is worth making some comments now about the comparison of our results with previous work,
in particular \cite{Arciniega:2018fxj}, where the series of curvature invariants was truncated at third order, finding an inflationary behavior. 
There, the general gravitational and cosmological properties of the model were studied but not the details of the inflationary process. We can notice that, in the scenario depicted in \cite{Arciniega:2018fxj}, the scale factor evolves as a power of time ($a(t)\sim t^{3/2}$), so we do not expect the slow-roll parameters to be within the observational constraints. In fact, by using for the cubic case the same method employed in this work, we
obtain the following, almost constant, values for the slow-roll parameters from 50 to 60 e-folds:
$\epsilon_1=2/3$, $\epsilon_2\sim 0$ and $\epsilon_3=8/3$ so $r=10.66$ and $n_s=-1.22$. In \cite{Edelstein:2020lgv, Edelstein:2020nhg} the authors introduce an extra field in order to have a proper inflationary evolution shared by the inflaton and the geometric contribution of the invariants to the third order.  Some other efforts have been performed in order to analyze the dynamics of the evolution when the
cubic contribution is taken into account  \cite{Cisterna:2018tgx, Quiros:2020eim, Quiros:2020uhr}. It is worth mentioning the analysis performed by \cite{Pookkillath:2020iqq} where it is shown that the evolution, for the cubic case in an anisotropic vacuum universe is not stable, opening new questions regarding the stability of the cubic model in a FLRW Universe.

\section{Conclusion}
\label{Conclusion}
We have found that the gravitational modified theory introduced in \cite{Arciniega:2018tnn} can achieve a viable inflationary epoch. Furthermore, all models considered in this work, but model 1, fulfill the slow-roll condition for $\epsilon_1$ and $\epsilon_3$, not so for $\epsilon_2$ where all models have values lower than the one reported by Planck 2018 for $N=50$ to $N=60$ at marginalized $68\%$ (we cannot compare with the confidence value at marginalized $95\%$ CL, since Planck does not report it). However, when the relationship between values in $\epsilon_{1,2}$ with respect to $\epsilon_3$ is analyzed, all models, but model 1, are in agreement with Planck inflationary constrictions 2018, and partially contained in contour region for the relation $\epsilon_1-\epsilon_2$. On another hand, it is interesting to notice that model 1 is the one that better fulfills the observational bound on $n_s$ at 1$\sigma$ but is far on the other observational parameters.
 
In summary, the models presented here are potentially successful inflationary scenarios.

Figure \ref{fig:nsvsr}, for which model 2 (which corresponds to model 1 in \cite{Arciniega:2018tnn} and goes like $\text{exp}(HL)^4$) is fully contained in the $95\%$ CL contour, is particularly noteworthing.

Although $\lambda$, related with the constant Lagrangian density coupling $\lambda_n$, has been ignored in the present work owing to the independence showed for the inflationary epoch considered here, it should be possible to constrain it with observations at late universe \cite{Arciniega2020}.

It is important to mention that, given that the constrictions presented by Planck 2018 are model dependent, the contours could change when the fiducial model is modified. In the case of the present theory, we could expect some changes regarding the values of  $\Omega_b h^2$ and $\Omega_c h^2$, nevertheless changes (if any) should be very small.

Finally, we want to remark that the theory of Geometric Inflation, in particular model 2, looks like a viable modified gravity for cosmology and is worthwhile to go further and explore it at some other cosmological and astrophysical scenarios.

\section*{Acknowledgments}
\noindent
GA wishes to acknowledge the postdoctoral fellowship from DGAPA-UNAM, LJ thanks the financial support from SNI (CONACyT) and the hospitality and fellowship to the Instituto Nacional de Investigaciones Nucleares (ININ). The authors aknowledge partial support from PAPIIT IN117817 and IN120620 .

\end{document}